\documentstyle[preprint,eqsecnum,aps,prb]{revtex}
\tighten
\begin{document}
\title{Nuclear spin-spin coupling in La$_{2-x}$Sr$_{x}$CuO$_{4}$ 
studied by stimulated echo decay}
\author{S.  Fujiyama\footnote{fujiyama@kodama.issp.u-tokyo.ac.jp}, M.  
Takigawa and Y.  Ueda }
\address{The Institute for Solid State Physics, The University of 
Tokyo, Roppongi, Minato-ku, \\ Tokyo 106-8666, Japan }
\author{T. Suzuki and N. Yamada} 
\address{Dept.  of Applied Physics and Chemistry, University of 
Electro-Communications, \\ Tokyo 182-8585, Japan }
\date{Apr. 20, 1999} 
\maketitle
\begin{abstract}
We have performed copper NQR experiments in high temperature 
superconductors YBa$_{2}$Cu$_{4}$O$_{8}$, YBa$_{2}$Cu$_{3}$O$_{7}$, 
and La$_{2-x}$Sr$_{x}$CuO$_{4}$ ($x$=0.12 and $x$=0.15), using the 
stimulated echo technique which utilizes the rf-pulse sequence 
$\pi/2-(\tau)-\pi/2-(T-\tau)-\pi/2$. The $\tau$ and $T$ dependences 
of the stimulated echo intensity is analyzed by a model that includes 
the spin-lattice relaxation process ($T_1$-process) and the 
fluctuating local field due to nuclear spin-spin coupling. The model 
gives quantitative account of the experimental results in 
YBa$_{2}$Cu$_{4}$O$_{8}$ and YBa$_{2}$Cu$_{3}$O$_{7}$, using the 
known values of $1/T_{1}$ and $1/T_{2G}$, the gaussian decay rate of 
the 
spin echo intensity. The same model applied to 
La$_{2-x}$Sr$_{x}$CuO$_{4}$ enables us to extract the value of 
$T_{2G}$. Our results indicate that $T_{1}T/T_{2G}$ is independent of 
temperature, implying that the dynamic exponent is one in 
La$_{2-x}$Sr$_{x}$CuO$_{4}$.
\end{abstract}
\narrowtext
\section{Introduction}
\label{sec:Intro}
Anomalous spin dynamics in high temperature superconductors continues 
to be a controversial issue\cite{Ginsberg,LosAlamos}. In all 
high-$T_{c}$ materials, the dynamic 
susceptibility $\chi(q, \omega )$ at low frequencies is peaked near 
or 
at the antiferromagnetic wave vector $Q=(\pi, \pi)$ above $T_{c}$. 
The nuclear spin-lattice relaxation rate ($1/T_{1}$) at the planar 
$^{63}$Cu sites measured by NMR/NQR has been used extensively to 
probe 
the low frequency spin fluctuations through the following relation 
\cite{Moriya1956}
\begin{equation}
\frac{1}{T_{1}T}=\frac{\gamma_{n}^{2}k_{B}}{\mu_{B}^{2}}\sum_{q}F_{\perp}(q)^{2}\chi '' 
(q, \omega_{n})/\omega_{n} ,
\label{eq:MoriyaT1}
\end{equation}
where $\gamma_{n}$ is the nuclear gyromagnetic ratio, $\omega_{n}$ is 
the nuclear resonance frequency, and $F_{\perp}(q)$ ($F_{\parallel}$) 
is the Fourier component of the hyperfine coupling perpendicular 
(parallel) to the crystalline $c$-axis, which is assumed to be the 
quantization axis of nuclear spins as is the case for NQR. It was 
found that $1/(T_{1}T)$ shows contrasting temperature dependence for 
different regions of doped hole concentration. For sufficiently high 
hole concentration, $1/(T_{1}T)$ keeps increasing as temperature is 
lowered down to $T_{c}$. In contrast, in many underdoped materials, 
where the hole concentration is smaller than the optimum value for 
highest $T_{c}$, $1/(T_{1}T)$ shows a broad maximum at a temperature 
$T_{s}$, which is much higher than $T_{c}$, and decrease steeply at 
lower temperatures. This suggests that a pseudo gap opens in the spin 
excitation spectrum near $(\pi, \pi)$ above $T_{c}$ 
\cite{Yasuoka1989}. 
As the hole concentration increases, $T_{s}$ decreases. The pseudo 
spin-gap phenomenon is observed in various underdoped materials, such 
as YBa$_{2}$Cu$_{3}$O$_{6.63}$\cite{Takigawa1991}, 
YBa$_{2}$Cu$_{4}$O$_{8}$\cite{Curro1997}, 
LaBa$_{2}$Cu$_{3}$O$_{7-\delta}$\cite{Goto1997} 
and HgBa$_{2}$Ca$_{n-1}$Cu$_{n}$O$_{2n+2+\delta} 
(n=1,2)$\cite{Itoh1996,Julien1996}.

However, one typical high-$T_{c}$ system, La$_{2-x}$Sr$_{x}$CuO$_{4}$ 
(LSCO) does not show clear pseudo spin-gap behavior\cite{Ohsugi1994}. Instead, recent 
precise measurements of $1/(T_{1}T)$ up to $800K$ 
\cite{Yasuoka1997,Fujiyama1997,Itoh1997} revealed a modest change of 
behavior at a temperature $T^{*}$. Above $T^{*}$, $1/(T_{1}T)$ 
follows a Curie-Weiss temperature dependence, $1/(T+\theta)$. Below 
$T^{*}$, $1/(T_{1}T)$ keeps increasing but is slightly suppressed 
from 
the extrapolation of the high temperature Curie-Weiss law. Dependence 
of $T^{*}$ on the hole concentration is similar to that of $T_{s}$ in 
other underdoped high-$T_{c}$ materials. At present, origin of this 
crossover temperature or the reason for the different behavior 
between 
LSCO and other high-$T_{c}$ materials is not well understood.

NQR experiments in LSCO are complicated due to large inhomogeneous 
line width of a few MHz. Since Cu spins are excited over a few 
hundred kHz by typical rf-pulses, only a small portion of the entire 
Cu spins contributes to the NQR signal at one frequency. In our 
previous work, it was found that $1/T_{1}$ measured by Cu-NQR is not 
uniform over the broad spectrum. The frequency dependence of 
$1/T_{1}$ 
is more pronounced in more underdoped materials\cite{Fujiyama1997}, 
indicating substantial inhomogeneity of the microscopic electronic 
state. One concern is that the measured value of $1/T_{1}$ might be 
influenced by the spectral diffusion process, in which Zeeman energy 
$\hbar\gamma_{n}I_{z}H$ of the excited nuclear spins diffuses in 
frequency over inhomogeneously broadened spectrum by the transverse 
component of nuclear spin-spin coupling of the form $aI_{i+}I_{j-}$. 
If this is the case, the measured value of $1/T_{1}$ would be larger 
than the true relaxation rate due to electronic spin fluctuations.

Another issue is that one cannot obtain the strength of nuclear 
spin-spin coupling from measurements of the spin echo decay rate for 
such a broad spectrum. In cuprates, Cu nuclear spins are coupled 
dominantly by the Ruderman-Kittel-Kasuya-Yoshida (RKKY) type indirect 
interaction mediated by electronic spin excitations,

\begin{equation}
{\cal H} = \hbar \sum_{<ij>} \left\{ a_{\parallel}^{ij} I_{iz}I_{jz}
+a_{\perp}^{ij}/2 \left( I_{i+}I_{j-} + I_{i-}I_{j+} \right) 
\right\}, 
\label{eq:IIHamiltonian}
\end{equation}
where the coupling constant is given in terms of the $q$-dependent 
static susceptibility as
\begin{equation}
a_{\alpha}^{ij} = \frac{1}{\hbar (g\mu_B)^2}\sum_{q}|F_{\alpha}(q)|^2 
\chi(q) \exp (-iqR_{ij}) .
\label{eq:RKKY}
\end{equation} 
The large anisotropy $F_{\parallel}(q)/F_{\perp}(q) \sim 6$ for $q 
\sim (\pi, \pi)$ makes $a_{\parallel}$ much larger than $a_{\perp}$. 
Then the Gaussian time constant of spin echo decay $1/T_{2G}$ in 
$^{63}$Cu NQR experiments is given as 
\cite{Pennington1991,Takigawa1994,Thelen1994}
\begin{equation} 
\left(\frac{1}{T_{2G}}\right)^{2}=\frac{0.69}{4}\sum_{j \neq i} 
(a_{\parallel}^{ij})^2 = 
\frac{0.69}{4\hbar^{2}}\frac{1}{(g\mu_{B})^{4}}\left[\sum_{q}F_{\parallel}
(q)^{4}\chi(q)^{2}-\left(\sum_{q}F_{\parallel}(q)^{2}\chi(q)\right)^{2}\right] , 
\label{eq:PenT2g}
\end{equation}
where 0.69 is the natural abundance of $^{63}$Cu nucleus.
Measurements of $1/T_{2G}$ have provided valuable information 
regarding the enhancement of the static susceptibility near the 
antiferromagnetic wave vector in many high-$T_{c}$ materials such as 
YBa$_{2}$Cu$_{3}$O$_{7-\delta }$(Y1237, $T_{c}\sim 
90K$)\cite{Pennington1989,Itoh1992}, YBa$_{2}$Cu$_{3}$O$_{6.62} 
$($T_{c}\sim 60K$)\cite{Takigawa1994}, and 
YBa$_{2}$Cu$_{4}$O$_{8}$(Y1248, $T_{c}\sim 80K$)\cite{Itoh1992}. 
However, Eq.\ (\ref{eq:PenT2g}) assumes that the nuclear spins over 
the entire spectrum are flipped by the $\pi$-pulse, a condition that 
is not satisfied in LSCO.

In this paper, we report the results of stimulated echo\cite{Hahn1950} decay 
measurements in YBa$_{2}$Cu$_{4}$O$_{8}$, YBa$_{2}$Cu$_{3}$O$_{7}$, 
and La$_{2-x}$Sr$_{x}$CuO$_{4}$ ($x$=0.12 and $x$=0.15), using NQR at 
zero magnetic field. Stimulated echo sequence utilizes three $\pi/2$ 
pulses separated in time by $\tau$ and $T-\tau$. The decay of 
stimulated echo intensity as a function of $T$ can be caused not only 
by the spin-lattice relaxation process but also by the spectral 
diffusion\cite{Weger} as well as the nuclear spin-spin 
coupling\cite{Klauder1962}. The 
dependence on $\tau \/$ and $T\/$ of the stimulated echo intensity is 
compared with the calculation based on a model which takes account of 
the spin-lattice relaxation process and the nuclear spin-spin 
coupling 
along the $c$-axis. The effects of spin-spin coupling perpendicular 
to the $c$-axis, which is much smaller than the coupling along the 
$c$-axis but responsible for the spectral diffusion, is neglected as 
we discuss in detail below. For YBa$_{2}$Cu$_{4}$O$_{8}$ and 
YBa$_{2}$Cu$_{3}$O$_{7}$, where the values of $1/T_{1}$ and 
$1/T_{2G}$ 
are known, the calculation shows good quantitative agreement with the 
experimental data. The same model applied to LSCO enables us to 
extract the value of $1/T_{2G}$. Our results indicate that 
$T_{1}T/T_{2G}$ is independent of temperature, implying that the 
dynamical exponent is one in LSCO.

\section{Sample preparations}
\label{sec:procedure}
We used polycrystalline materials synthesized by the usual solid 
state reaction. For Y1248 material, we used hot-isostatic-pressing 
(HIP) technique. 
All samples were confirmed to be single phase by powder X-ray 
diffraction measurements. 
The superconducting transition temperatures ($T_{c}$) are 90K (Y1237), 
80K (Y1248), 38K (LSCO $x=0.15$) and 32K (LSCO $x=0.12$), 
respectively. 

\section{Stimulated echo method}
\label{sec:stecho}
We briefly discuss the stimulated echo method in comparison with the 
ordinary spin echo and inversion recovery methods. In ordinary spin 
echo experiments, two rf-pulses ($\pi/2$ and $\pi$ pulses) are 
applied 
with time separation $\tau$. The $\tau$ dependence of spin echo 
intensity is described by two parameters, $1/T_{2L}$ and $1/T_{2G}$ 
as\cite{Pennington1989},
\begin{equation}
M(2\tau )=M_{0}\exp \left(-\frac{2\tau }{T_{2L}}\right)\exp 
\left(-\frac{(2\tau )^{2}}{2T_{2G}^{2}}\right) .
\label{eq:echodecay}
\end{equation}
The Lorenzian component $1/T_{2L}$ represents the contribution from 
the spin-lattice relaxation process\cite{Slichter}. The Gaussian 
component $1/T_{2G}$ is due to nuclear spin-spin coupling 
and given by Eq.\ (\ref{eq:PenT2g}).

The pulse sequence for the stimulated echo experiments are described 
as $(\pi/2 - \tau - \pi/2 - (T-\tau)-\pi/2)$, which is shown in Fig.\ 
\ref{fig:stmlecho}(b) and compared with the conventional inversion 
recovery sequence ($\pi - T - \pi/2 - \tau - \pi$) in Fig.\ 
\ref{fig:stmlecho}(a). The figures below the time chart of rf-pulses 
show the corresponding state of nuclear spins in the rotating frame. 
In the inversion recovery sequence (a), the nuclear magnetization 
along the $z$-direction initially at the thermal equilibrium is 
inverted by the first $\pi$ pulse. The $z$-component of the 
magnetization $M_z$, which then begins to recover toward the thermal 
equilibrium, is measured by the intensity of the spin-echo signal 
formed by the second and the third pulses. In case of NQR of 
$^{63}$Cu nuclei ($I=3/2$), the recovery of spin echo intensity is 
described by an exponential function if the effect of spectral 
diffusion is neglected,
\begin{equation}
M(T)=M_{0}\left[ 1 - 2 \exp \left( -\frac{3T}{T_{1}}\right) \right] ,
\label{eq:T1recovery}
\end{equation}
where $T_{1}$ is the time constant for high field NMR experiments.

In the stimulated echo sequence (b), the initial magnetization is 
flipped into the $y$-direction by the first $\pi$/2 pulse. Then it 
dephases in the $xy$-plane by the inhomogeneous distribution of the 
resonance frequency, as well as by the nuclear spin-spin coupling. 
The magnetization spread in the $xy$-plane is flipped into the 
$xz$-plane by the second $\pi/2$ pulse. The value of $M_{z}$ 
immediately after the second pulse is given by the ensemble average of 
$-\cos\phi(\tau)$, where $\phi(\tau)=\gamma_n \int_{0}^{\tau} h(t)dt$ 
is the accumulated phase in the $xy$-plane of individual spins subject 
to the local field $h(t)$ during the period between the first and 
second pulses. If the local field distribution is dominated by the 
static inhomogeneity of the resonance frequency, $M_{z}$ should be 
oscillating as $-\cos(\tau \omega)$ as a function of frequency over 
the spectrum as shown in Fig.\ \ref{fig:accumulated}. If $M_{z}$ does 
not change between the second and the third pulse, the oscillating 
magnetization distribution along the $z$-direction is flipped into the 
$y$-direction by the third pulse, which then refocuses along the 
$-y$-direction at $T+\tau$, forming the stimulated echo. The echo 
intensity is given by the ensemble average of $(1/2)\cos\phi(T+\tau)$, 
where
\begin{equation}
\phi(T+\tau) 
=\gamma_{n}\int_{0}^{\tau}h(t)dt-\gamma_{n}\int_{T}^{T+\tau } h(t)dt 
\label{eq:accumulated}
\end{equation}
is the accumulated phase at $T+\tau$. The accumulated phase is zero if $h(t)$ is 
static.

Temporal variation of $M_{z}$ or $h(t)$ causes the intensity of 
stimulated echo to decay as a function of $T$. Change of $M_{z}$ can 
be caused either by the spin-lattice relaxation process 
($T_{1}$-process) or by the spectral diffusion. By the 
$T_{1}$-process $M_{z}$ recovers towards the uniform thermal 
equilibrium magnetization. The stimulated echo decay by the 
$T_{1}$-process is simply given by $\exp (-3T/T_{1})$ similar to the 
inversion recovery.

The spectral diffusion can be caused by the transverse component of 
the nuclear spin-spin coupling, the second term of Eq.\ 
(\ref{eq:IIHamiltonian}), which describes the mutual flip of two 
nuclear spins. We discuss this in more detail in the next section. 
The oscillating distribution of $M_{z}$ will be averaged out by the 
spectral diffusion, leading to reduction of the stimulated echo 
intensity.

Stimulated echo decay is also caused by the temporal fluctuation of 
$h(t)$, since the two terms in the accumulated phase Eq.\ 
(\ref{eq:accumulated}) then do not cancel exactly. In our NQR 
experiments, distribution of the local field is largely due to 
inhomogeneity of the electric field gradient, which is static. 
However, the longitudinal component of the spin-spin coupling, the 
first term of Eq.\ (\ref{eq:IIHamiltonian}), gives rise to an 
additional source of the local field. This local field at $i$-th site 
is given as
\begin{equation}
h_{i}(t) = \frac{1}{\gamma_{n}}\sum_{j \neq 
i}a_{\parallel}^{ij}I_{jz}(t) ,
\label{eq:localf1}
\end{equation}
which fluctuates via the $T_{1}$-process.
 
\section{Experimental results and analysis}
\label{sec:results}
The stimulated echo intensity $M$ was measured as a function of $T$ 
for several different values of $\tau$. The experimental results are 
shown in Figs. \ref{fig:rawY1248}, \ref{fig:rawY1237} for Y-based 
compounds and Fig. \ref{fig:rawLSCO} for LSCO $x=0.12$ sample. 
Generally, the echo decay curve $M(T)$ deviates from a single 
exponential function $\exp (-3T/T_{1})$. The deviation is not clearly 
noticed for small $\tau$ but becomes significant for larger $\tau$. 
The decay rate $d\ln M(T)/dT$ shows strong $\tau$ dependence for small 
$T$, where it is larger for larger $\tau$. In contrast, the decay 
rate depends less on $\tau$ for larger $T$ region. How strongly the 
decay curve changes with $\tau$ depends on temperature as well as 
materials. As shown in Figs. \ref{fig:rawY1248}, \ref{fig:rawY1237}, 
and \ref{fig:rawLSCO}, the $\tau$ dependence of stimulated echo decay 
curves become negligible at higher temperatures. At lower 
temperatures, the $\tau$ dependence of decay curves is most 
significant in Y1248 sample.

As mentioned in the previous section, we have to consider three 
distinct processes for the stimulated echo decay, namely, the 
$T_{1}$-process, spectral diffusion, and local field fluctuations. Of 
these, contribution from the $T_{1}$-process is easily separated by 
dividing the experimental data of stimulated echo decay by $\exp 
(-3T/T_{1})$.

Portis has analyzed the spectral diffusion process in electron 
paramagnetic resonance of F-centers, where diffusion occurs via the 
dipolar interaction between electron spins within the spectra 
broadened by the hyperfine interaction with surrounding nuclear spins \cite{Portis1956}. We apply this analysis to the present problem. 
First, the inverse life time of the $z$-component of a nuclear spin 
due to mutual spin flip of $I_{i}$ and a nearby spin $I_{j}$ described 
by the second term of Eq.\ (\ref{eq:IIHamiltonian}) is given by 
\cite{Portis1956}
\begin{equation}
\frac{1}{\tau_{s}}=0.69\times \frac{\pi}{4} f(\omega)
\sum_{j}(a_{\perp}^{ij})^2 ,
\label{eq:avtransition}
\end{equation}
where $f(\omega)$ is the NQR spectral shape function normalized so 
that $\int f(\omega) d\omega = 1$ and we assume that the resonance 
frequencies of two neighboring nuclear spins are uncorrelated. 
Although the total Zeeman energy has to be conserved, the uncertainty 
principle allows the resonance frequencies of $I_{i}$ and $I_{j}$ to 
be different by an amount comparable to $1/\tau_{s}$. Thus the mutual 
spin flip process describes a one dimensional random walk of magnetic 
particles along the frequency axis, where both the hopping rate and 
the hopping distance are given by $1/\tau_{s}$. Therefore, the time 
evolution of the macroscopic magnetization distribution is given by a 
diffusion equation with the diffusion constant 
$D_{\omega}=1/\tau_{s}^{3}$. We can estimate $1/\tau_{s}$ as follows. 
From 
Eq.\ (\ref{eq:PenT2g}),
\begin{equation}
\frac{1}{\tau_{s}}=\pi f(\omega) \left( \frac{1}{T_{2G}} \right)^2 
\frac{\sum_{i \neq j} (a_{\perp}^{ij})^2}{\sum_{i \neq 
j}(a_{\parallel}^{ij})^2} ,
\label{eq:avtransition2}
\end{equation}
where $\sum_{i \neq j}(a_{\perp}^{ij})^2 / \sum_{i \neq 
j}(a_{\parallel}^{ij})^2 \sim \{F_{\perp}(Q) / F_{\parallel}(Q)\}^4 
\sim 1/40$ in high-$T_{c}$ cuprates. Among the materials we studied, 
$\tau_{s}$ should be shortest in Y1248 because of small NQR line width 
of 200kHz ($f(\omega)=(4\pi \times 10^{5})^{-1}$) and relatively large 
$1/T_{2G} \sim 3 \times 10^{4}$ sec$^{-1}$ near $T_{c}$. From these 
numbers, $1/\tau_{s}$ is estimated to be 56 sec$^{-1}$.

In principle, such spectral diffusion could average out the 
oscillation of the magnetization distribution along the frequency axis 
with the period $2\pi/\tau$ in the stimulated echo experiments (Fig.\ 
\ref{fig:accumulated}). The characteristic time for this to occur is 
given by $(\pi/\tau)^2/D_{\omega}$. Even for the longest value of 
$\tau=5 \times 10^{-5}$sec in our experiments, this characteristic 
time is of the order of $10^{4}$sec, which is many orders of magnitude 
larger than the range of $T$ in our experiments (less than 
$10^{-2}$sec). Therefore, we conclude that the effect of the spectral 
diffusion is negligible.

We now consider the effect of fluctuations of the local field. 
Recchia {\it et al.} have considered the same problem for the 
ordinary 
spin echo decay \cite{Recchia}. They assumed a Gaussian distribution 
for the accumulated phase and derived an analytic expression for the 
spin echo decay in terms of the correlation function of the local 
field. Curro {\it et al.} applied this results to the case where the 
local field is given by Eq.\ (\ref{eq:localf1}) and $I_{z}$ of a 
neighboring nuclear spin is fluctuating via the $T_{1}$ process 
\cite{Curro1997}. This approach can be applied to the stimulated echo 
decay with only minor modification. If we assume a Gaussian 
distribution for the accumulated phase defined by Eq.\ 
(\ref{eq:accumulated}), the stimulated echo intensity is given by
\begin{equation}
\langle \frac{1}{2}\cos\phi(T+\tau)\rangle=\frac{1}{2}\exp\left(-\frac{\langle
\phi(T+\tau)^2\rangle}{2}\right) . 
\label{eq:gaussap}
\end{equation}
As discussed in the appendix, $\langle\phi(T+\tau)^2\rangle$ is the 
ensemble average of the contribution from individual neighboring 
nuclear spins. 
Each of them has to be distinguished whether it is a like nucleus, 
which is on resonance, or an unlike nucleus, which is off resonance. 
Thus if there are $n$ types of unlike nuclei, we can write
\begin{equation}
\langle\phi(T+\tau)^2\rangle = 
P_{0}\langle\phi_{like}(T+\tau)^2\rangle + \sum_{i=1}^{n} 
P_{i}\langle\phi_{unlike,i}(T+\tau)^2\rangle ,
\label{eq:decomp}
\end{equation}
where $P_{0}$ and $P_{i}$ are the abundance of the like nuclei and 
$i$-th unlike nuclei, respectively. In the appendix, 
$\langle\phi_{like}(T+\tau)^2\rangle$ and 
$\langle\phi_{unlike,i}(T+\tau)^2\rangle$ are calculated as functions 
of $T_{2G}$ of $^{63}$Cu nuclei as defined in Eq.\ (\ref{eq:PenT2g}) 
and $T_{1}$'s of like and unlike nuclei. In Fig. \ref{fig:phiplot}, 
$T$ dependence of $\langle\phi_{like}(T+\tau)^2\rangle$ and 
$\langle\phi_{unlike}(T+\tau)^2\rangle$ are plotted for different 
values of $T_{2G}/T_{1}$ and $\tau/T_{1}$.

To summarize our analysis, we expect that the $T$ and $\tau$ 
dependence of the stimulated echo intensity $M(T, \tau )$ is given by 
the product of the two decay process, i.e. the $T_{1}$-process and 
local field fluctuations,
\begin{equation}
M(T, \tau ) = M_{0} \exp \left( -\frac{3T}{T_{1}}\right) \exp \left( 
- 
\frac{\langle\phi(T+\tau)^2\rangle}{2}\right) .
\label{eq:totdecay}
\end{equation}

\section{Discussions}
\label{sec:fitting}
In this section we compare the experimental results with the 
calculation of Eq.\ (\ref{eq:totdecay}). In order to examine the 
validity of our approach to analyze the local field fluctuations, we 
separate the trivial $T_{1}$-process. In the following, we compare 
the stimulated echo intensity divided by $\exp ( -3T/T_{1})$, which we 
call the corrected intensity $M_{corr}(T, \tau )$, with the 
calculated results of $\exp (-\langle\phi(T+\tau)^2\rangle / 2)$.
The results of comparison are shown in Figs. \ref{fig:Y1248}, 
\ref{fig:LSCO}, and \ref{fig:LSCOfitTrial}.

\subsection{Y-based compounds}
We first discuss the results on underdoped YBa$_{2}$Cu$_{4}$O$_{8}$ 
and optimally doped YBa$_{2}$Cu$_{3}$O$_{7-\delta }$ ($T_{c}$=90K). 
The full width at half maximum of $^{63}$Cu-NQR spectra is about 200 
kHz in these samples. Since almost all $^{63}$Cu nuclear spins are 
excited by rf-pulses for such a narrow spectrum, all $^{63}$Cu spins are 
like nuclei and $^{65}$Cu spins are the only unlike nuclei. Thus we set 
$P_{0}=0.69$ and $P_{1}=0.31$ in Eq.\ (\ref{eq:decomp}). In Fig.\ 
\ref{fig:Y1248}, the data for $M_{corr}$ and the calculated 
stimulated 
echo decay curves are compared in Fig. \ref{fig:Y1248} for Y1248 and 
Y1237 compounds. For calculation, we used the value of $T_{1}$ 
obtained from NQR inversion recovery measurements and the data of 
$T_{2G}$ obtained by spin echo decay measurements \cite{Itoh1992}. 
The overall magnitude $M_{0}$ is the only adjustable parameter in 
this 
comparison. The agreement between the experimental data and 
calculation is quite good. The calculation reproduces the observed 
change of behavior for different values of $\tau$, in particular, the 
non-exponential decay for large values of $\tau$ is well explained by 
our model. Thus we believe that our model is valid for these 
materials.

\subsection{La$_{2-x}$Sr$_{x}$CuO$_{4}$}
As we mentioned before, the Cu-NQR spectra of LSCO show significant 
inhomogeneous broadening, whose full width at half maximum (FWHM) is 
about 2MHz. Thus it is not possible to obtain $1/T_{2G}$ as defined 
in Eq.\ (\ref{eq:PenT2g}). On the other hand, Walstedt {\em et al.\/} 
measured the spin echo decay rate using the $I_{z}=1/2\leftrightarrow -1/2$ 
center line of the $^{63}$Cu NMR in the oriented powder sample in high 
magnetic field\cite{Walstedt1996}.  The FWHM of the NMR center line in 
LSCO is typically about 500Oe.  Although this is about
4 times narrower than that of Cu 
NQR, the whole region of the spectra can not be excited by rf-pulses. 
Moreover, the NMR spectra includes both the broad background from 
unoriented portion of the sample and the resonance from the so called 
{\em B-\/}site, which may affect the spin echo decay measurements. Thus 
it would be important to obtain $1/T_{2G}$ by a different method.

Here we assume that our model for the stimulated echo decay is valid 
also for LSCO and extract the value of $1/T_{2G}$ by fitting the data 
of $M_{corr}$ to $\exp (-\langle\phi(T+\tau)^2\rangle / 2)$ with 
$T_{2G}$ as a fitting parameter. The experimental data and fitted 
curves for LSCO are shown in Fig.\ \ref{fig:LSCO}. We used the value 
of $T_{1}$ determined from the inversion recovery technique.

There is minor ambiguity in this procedure. Since the $^{63}$Cu NQR 
spectrum is so broad, only a fraction of $^{63}$Cu spins is considered 
to be like nuclei. However, this is not a serious problem, since 
$\langle\phi_{like}(T+\tau)^2\rangle$ and 
$\langle\phi_{unlike}(T+\tau)^2\rangle$ are nearly identical for the
range of parameters of our interest if the values of $1/T_{1}$ and 
$\gamma_{n}$ are the same, as shown in Figs. \ref{fig:phiplot} and 
\ref{fig:LSCOfitTrial}. Based on the ratio between the NQR line width 
and the magnitude of the rf magnetic field of our experiments ($\sim$ 
200kHz), we chose rather arbitrarily that 10\% of $^{63}$Cu are like 
nuclei and 90\% are unlike nuclei. The obtained value of $1/T_{2G}$ 
hardly depends on these numbers. We show in Fig. 
\ref{fig:LSCOfitTrial} two fits obtained by assuming that 100\% of 
$^{63}$Cu are like or unlike nuclei. The fitted values of $T_{2G}$ 
differ only by 6\%. Our analysis to determine $T_{2G}$ is limited below 150K for $x=0.15$ 
and below 250K for $x=0.12$. At higher temperatures, where 
$1/T_{2G}$ gets smaller, $T$ dependence of $M_{corr}$ becomes too 
weak to determine $1/T_{2G}$ reliably. The values of $T_{2G}$ thus 
derived are plotted against temperature for two samples, $x=0.12$ and 
$x=0.15$ in Fig. \ref{fig:T2gLSCO}. $1/T_{2G}$ increase steeply as 
temperature decreases, similar to other high-$T_{c}$ 
materials\cite{Pennington1989,Takigawa1994,Itoh1992}. The results 
obtained by Walstedt {\em et al.\/} for $x=0.15$ between $T=100K$ and 
300K using high field NMR are also shown by the dashed line, after 
multiplied by $\sqrt{2}$ in order to take account of the difference 
between NQR and NMR measurements\cite{Walstedt1995,Walstedt1996}.

Since the antiferromagnetic correlation is quite strong in high-$T_{c}$ 
cuprates, we generally expect that the dynamic spin correlations obey 
some kind of scaling relation, which relates temperature dependence of
various magnetic quantities to those of the antiferromagnetic 
correlation length $\xi$ and the characteristic energy of the 
antiferromagnetic spin fluctuations $\omega_{sf}$. These two 
quantities are related by the dynamical exponent $z$ as $\omega_{sf} 
\propto \xi^{-z}$. The data of $1/T_{2G}$ and $1/T_{1}T$ have been 
used to extract the value of $z$ in various high-$T_{c}$ cuprates. It 
has been shown that if $\chi(Q) \propto \xi^{2}$,
\begin{equation}
\frac{1}{ T_{1} T}\propto \omega_{sf}^{-1} , \frac{1}{ T_{2G}}
\propto \xi .
\label{eq: T1T2g}
\end{equation} 
Therefore, we expect $T_{1}T/T_{2G}^z$ to be temperature independent 
\cite{Sokol1993}.  While $z=1$ is expected at the quantum critical 
point (the boundary between N\'{e}el ordered states and disordered 
states) in two dimensional quantum spin system\cite{Chubukov1993}, 
antiferromagnetic spin fluctuations in itinerant electron systems is 
generally believed to be described by $z=2$.

Experimentally, the underdoped materials of the Y-based system, 
YBa$_{2}$Cu$_{3}$O$_{6.63}$ and YBa$_{2}$Cu$_{4}$O$_{8}$ shows the 
$z$=1 behavior above $T_{s}$. Early data on the optimally doped 
YBa$_{2}$Cu$_{3}$O$_{6.9}$ are consistent with $z$=2 
\cite{Sokol1993,Imai1993}. However, recent $^{17}$O-NMR experiments 
by Keren {\it et al.} pointed out importance of fluctuations of Cu 
nuclear spins due to both spin-lattice and mutual flip process in the 
analysis of the spin echo decay data \cite{Keren1997}. They obtained 
$z$=1 after correcting these effects \cite{Keren1997}.

In fig.\ref{fig:z}, we plot the ratio $T_{1}T/T_{2G}^{z}$ for $x$=0.12 
and 0.15 against temperature with both $z$=1 and $z$=2.  Apparently, 
the plot for $z$=1 is more temperature independent than the plot for 
$z=2$.  Thus our results suggest the quantum critical scaling for the 
magnetic excitations in underdoped LSCO.  The results obtained by 
Walstedt {\em et al.\/} for $x=0.15$ between 100K and 300K are also 
consistent with $z=1$\cite{Walstedt1996}.  To summarize our results 
on LSCO, consistent account of the stimulated echo decay data is given 
by our model using the values of $1/T_{1}$ measured by the inversion 
recovery technique, which show no clear signature of pseudo spin-gap 
above $T_{c}$.  The quantum critical scaling we found for LSCO is 
also consistent with the lack of spin-gap, since such scaling breaks 
down in many of the underdoped cuprates below $T_{s}$.  Why the spin 
gap is absent in LSCO is not fully understood.  However, we speculate 
that it may be closely related the instability toward the simultaneous 
spin and charge ordering found experimentally in 
LSCO\cite{Tranquada1995}.

\section{Conclusion}
We have applied the stimulated echo technique to the high-$T_{c}$ 
superconductors by $^{63}$Cu NQR. Our model assumes that the 
stimulated echo decay is caused by two dominant processes, the 
longitudinal $T_{1}$ relaxations and fluctuation of local field 
produced by longitudinal component of the nuclear spin-spin coupling. 
We argue that effects of spectral diffusion due to transverse 
component of the spin-spin coupling is negligible. We confirm that 
our model indeed gives good quantitative account of the experimental 
data for Y-based materials. We used the same model to deduce the 
value of $1/T_{2G}$ from the data of stimulated echo decay in LSCO. 
We found that $1/T_{2G}$ increases monotonously with decreasing 
temperature. The results show that the ratio $T_{1}T/T_{2G}^{z}$ with 
$z$=1 is approximately independent of temperature, suggesting the 
quantum critical scaling in underdoped LSCO.

\acknowledgments
We are grateful to Prof.  H.  Yasuoka and Dr.  Y.  Itoh for suggesting 
stimulated echo experiments on high-$T_{c}$ superconductors.  One of 
the authors (S.  F.) thanks to S.  Onoda for stimulating discussions.  
This work is supported by the Grant in Aid of the Ministry of 
Education, Sports and Culture.  One of the authors (S.  F.) is 
supported by JSPS Research Fellowship for Young Scientists.

\appendix
\section*{}
We calculate the second moment of the accumulated phase defined in 
Eq.\ (\ref{eq:accumulated}), taking account of the fluctuation of the 
local field defined in Eq.\ (\ref{eq:localf1}) due to the $T_{1}$ 
process. We follow the approach formulated by Curro {\it et al.} 
\cite{Curro1997}. Since the values of $I_{jz}$ is jumping among the 
four eigenvalues $\pm 3/2$, $\pm 1/2$, the local field is written as
\begin{equation}
h_{i}(t) = \frac{1}{\gamma_{n}}\sum_{j \neq 
i}\sum_{m}a_{\parallel}^{ij}mp_{jm}(t) ,
\label{eq:localf2}
\end{equation}
where $p_{jm}$ is a probability that $I_{jz}$ takes the eigenvalue 
$m$. From Eqs.\ (\ref{eq:localf2}) and (\ref{eq:accumulated}), the 
second moment of the accumulated phase is given as
\begin{equation}
\langle \phi ^{2}(T+\tau ) \rangle =\left( \int_{0}^{\tau 
}\int_{0}^{\tau }-\int_{0}^{\tau }\int_{T}^{T+\tau }-\int_{T}^{T+\tau 
}\int_{0}^{\tau }+\int_{T}^{T+\tau }\int_{T}^{T+\tau } \right) 
\sum_{j \neq i} 
(a_{\parallel}^{ij})^2 \sum_{m,m'}mm'\langle 
p_{jm}(t)p_{jm'}(t')\rangle dtdt' ,
\label{eq:ensemble}
\end{equation}
where we have assumed no correlation between different nuclear spins. 

The time evolution of $p_{m}(t)$ is governed by the rate equation,
\begin{equation}
\frac{dp_{m}(t)}{dt}=\sum_{n}W_{mn}p_{n}(t)
\label{eq:mastereq}
\end{equation}

\begin{equation}
{\bf W}= \frac{1}{2T_{1}}
\times 
\left( 
\begin{array}{rrrr}
-3 & 3 & 0 & 0 \\
3 & -7 & 4 & 0 \\
0 & 4 & -7 & 3 \\
0 & 0 & 3 & -3
\end{array}
\right)
\end{equation}
We can write $\langle p_{m}(t) p_{m'}(t')\rangle = P_{mm'}(|t-t'|)/4$, 
where $P_{mm'}(t)$ is the probability that if a nuclear spin is in the 
state $m$ at the time $t$ under the condition that it was in the 
state $m'$ at $t=0$.  
$P_{mm'}(t)$ is equal to $p_{m}(t)$ obtained by solving the rate 
equation, Eq.\ (\ref{eq:mastereq}), with the initial condition that 
$p_{m'}=1$ at $t=0$.  For instance,
\begin{equation}
P_{\frac{3}{2}\frac{3}{2}}(t)=\frac{1}{4}+\frac{9}{20}
\exp(-\frac{t}{T_{1}}) 
+\frac{1}{4}\exp(-\frac{3t}{T_{1}})+\frac{1}{20}\exp
(-\frac{6t}{T_{1}}) .
\label{eq:Peg}
\end{equation}

We now have to consider the effects of three $\pi/2$ pulses , which 
redistribute the populations of like nuclear spins, which are on 
resonance. The rate equation, Eq.\ (\ref{eq:mastereq}), should be 
combined with the boundary conditions at $t=\tau$,
\begin{equation}
p_{3/2}(\tau^{+})=p_{1/2}(\tau^{+})=
\frac{1}{2}[p_{3/2}(\tau^{-})+p_{1/2}(\tau^{-})] .
\label{eq:boundery1}
\end{equation}
\begin{equation}
p_{-3/2}(\tau^{+})=p_{-1/2}(\tau^{+})=
\frac{1}{2}[p_{-3/2}(\tau^{-})+p_{-1/2}(\tau^{-})] .
\label{eq:boundery2}
\end{equation}
and the similar conditions at $t=T+\tau$. The rf-pulses have no 
effect for unlike nuclear spins, which are off resonance.

Eq.\ (\ref{eq:ensemble}) shows that $\langle\phi(T+\tau)^2\rangle$ is 
the sum of the contribution from individual neighboring nuclear 
spins. 
Therefore it is written as Eq.\ (\ref{eq:decomp}) according to the 
distribution of different types of like and unlike nuclei. We define 
$\phi_{like}(T+\tau)$ to be the accumulated phase for the 
hypothetical 
case where like nuclei occupy all the neighboring sites. Likewise 
$\phi_{unlike}(T+\tau)$ is defined to be the accumulated phase when 
all neighboring sites are occupied by $i$-th unlike nuclei. They are 
calculated as,

\begin{eqnarray}
\langle\phi_{like}^{2}(T+\tau)\rangle&=&\frac{2}{0.69}
\left(\frac{2T_{1}}{T_{2G}}\right)^{2}\left[\frac{5}{4}\left\{\frac{2\tau}{T_{1}}
-2+2\exp\left(-\frac{\tau}{T_{1}}\right)\right\}@\right. 
 \nonumber \\ & & \left. 
-\left\{\frac{4}{5}\exp\left(-\frac{T-\tau}{T_{1}}\right)
+\frac{1}{5}\exp\left(-\frac{6(T-\tau)}{T_{1}}\right)\right\}\left\{
1-\exp\left(-\frac{\tau}{T_{1}}\right)\right\}^{2}\right]
\label{eq:likephi}
\end{eqnarray}
and 
\begin{eqnarray}
\langle\phi_{unlike,i}^{2}(T+\tau)\rangle&=&\frac{5}{2 \times 0.69} 
\left(\frac{2T_{1i}}{T_{2G}}\right)^{2}\left(\frac{\gamma_{i}}
{\gamma_{0}}\right)^{2}\left[\frac{2\tau}{T_{1i}}+2\exp\left(-\frac{\tau}{T_{1i}}\right)
-2+2\exp\left(-\frac{T}{T_{1i}}\right) 
\right. \nonumber \\ & & \left. 
-\exp\left(-\frac{T+\tau}{T_{1i}}\right)-\exp\left(-\frac{T-\tau}{T_{1i}}\right)\right] .
\label{eq:unlikephi}
\end{eqnarray}

Here $1/T_{2G}$ is the $^{63}$Cu NQR spin echo decay rate as defined 
by Eq.\ (\ref{eq:PenT2g}), $1/T_{10}$ and $1/T_{1i}$ is the 
spin-lattice relaxation rate of like and $i$-th unlike nuclei, and 
$\gamma_{0}$ and $\gamma_{i}$ is the gyromagnetic ratio of like 
($^{63}$Cu) and $i$-th unlike nuclei respectively. The result for 
$\langle \phi_{unlike,i}^{2}(T+\tau))\rangle$ has obtained by Recchia 
{\it et al}. previously \cite{Recchia1996}.

\begin{figure}
\caption{Time charts of three rf-pulses and corresponding states of 
nuclear spins in the rotating frame for (a) Inversion recovery and (b) 
stimulated echo decay sequences.}
\label{fig:stmlecho}
\end{figure}

\begin{figure}
\caption{Distribution of $M_{z}$ in the 
inhomogeneously broadened 
spectrum in stimulated echo experiments after the second pulse. Typical 
values of the  $t_{w}$ ($\pi /2$ pulse width) and $\tau$ are 1.5$\mu s$ and 30$\mu 
s$.}
\label{fig:accumulated}
\end{figure}

\begin{figure}
\caption{Stimulated echo intensity is plotted against $T$ for various 
values of $\tau $ for YBa$_{2}$Cu$_{4}$O$_{8}$ at 
100K.}
\label{fig:rawY1248}
\end{figure}

\begin{figure}
\caption{Stimulated echo intensity for  YBa$_{2}$Cu$_{3}$O$_{7}$ at 100K.}
\label{fig:rawY1237}
\end{figure}

\begin{figure}
\caption{Stimulated echo intensity for La$_{2-x}$Sr$_{x}$CuO$_{4}$ 
$x=0.12$ sample at 100K (above) and at 300K (below).}
\label{fig:rawLSCO}
\end{figure}

\begin{figure}
\caption{The second moment of the accumulated phase $\langle \phi^{2} \rangle$ 
as a function of $T/T_{1}$ for different values of $T_{2G}/T_{1}$ 
and $\tau/T_{1}$ for (a)like-spin coupling and (b)unlike-spin coupling. }
\label{fig:phiplot}
\end{figure}

\begin{figure}
\caption{The experimental stimulated echo decay,  $M_{corr}$ (dots) 
is compared with the calculations (lines) for Y1248 and 
Y1237 at 100K. We used the value of parameter, $T_{2G}$ given in 
ref.18.
}
\label{fig:Y1248}
\end{figure}

\begin{figure}
\caption{The experimental stimulated echo decay, $M_{corr}$ (dots) and fitted 
curves based on Eq. 4.5.
}
\label{fig:LSCO}
\end{figure}

\begin{figure}
\caption{The experimental data of $M_{corr}$ for $x=0.15$ at 100K 
(dots) are compared with two fitted curves, assuming 100\% like spin 
coupling (solid curve) or 100\% unlike spin coupling (dotted curve).
The fitted values of $T_{2G}$ are 
25.0[$\mu s$] for like-spin coupling and 26.7[$\mu s$] 
for unlike-spin coupling.}
\label{fig:LSCOfitTrial}
\end{figure}

\begin{figure}
\caption{Temperature dependence of $1/T_{2G}$ for LSCO $x=0.12$ 
and $x=0.15$ samples. The dotted line shows the results by Walstedt 
{\em et al.\/} for $x=0.15$ obtained from the spin echo decay 
measurements using high field NMR.
}
\label{fig:T2gLSCO}
\end{figure}

\begin{figure}
\caption{The plot of $T_{1}T/T_{2G}^{z}$ (z=1 or 2) in LSCO $x=0.12$ 
and 0.15 against temperature.}
\label{fig:z}
\end{figure}


\begin{references}
\bibitem{Ginsberg}As reviews, {\em Physical Properties of High 
Temperature 
Superconductors 1-3\/}, ed. by D. M. Ginsberg (World Scientific, 
Singapore, 1989, 1990, 1993).
\bibitem{LosAlamos}As a review, {\em Strongly Correlated Electronic 
Materials\/}, ed. by K. S. Bedell {\em et al.\/} (Addison-Wesley, Reading, 
1994).
\bibitem{Moriya1956}T. Moriya, Prog. Theor. Phys. {\bf 16}, 641 
(1956).
\bibitem{Yasuoka1989}H. Yasuoka, T. Imai and T. Shimizu, {\em Strong 
Correlation and Superconductivity\/}, eds. by H. Fukuyama, S. Maekawa 
and A. P. Malozemoff, (Springer-Verlag, 1989).
\bibitem{Takigawa1991}M. Takigawa, A. P. Reyes, P. C. Hammel, J. D. 
Thompson, R. H. Heffner, Z. Fisk and K. C. Ott, Phys. Rev. B {\bf 
43}, 247 (1991). 
\bibitem{Curro1997}N. J. Curro, T. Imai, C. P. Slichter and B. 
Dabrowski, Phys. Rev. B {\bf 56}, 877 (1997), R. L. Corey, N. J. 
Curro, K. O'Hara, T. Imai, C. P. Slichter, K. Yoshimura, M. Katoh and 
K. Kosuge, Phys. Rev. B {\bf 53}, 5907 (1996).
\bibitem{Goto1997}A. Goto, H. Yasuoka, K. Otzschi and Y. Ueda, Phys. 
Rev. B {\bf 55}, 12736 (1997).
\bibitem{Itoh1996}Y. Itoh, T. Machi, A. Fukuoka, K. Tanabe and H. 
Yasuoka, J. Phys. Soc. Jpn. {\bf 65}, 3751 (1996), {\em ibid.\/} {\bf 
67}, 312 (1998).
\bibitem{Julien1996}M. H. Julien, P. Caretta, M. Horvati\'{c}, C. 
Berthier, P. S\'{e}gransan, A. Carrington and D. Colson, Phys. Rev. 
Lett. 
{\bf 76}, 4238 (1996).
\bibitem{Ohsugi1994}S. Ohsugi, Y. Kitaoka, K. Ishida, G. Zheng and K. 
Asayama, J. Phys. Soc. Jpn. {\bf 63}, 700 (1994), S. Ohsugi, Y. 
Kitaoka, K. Ishida and K. Asayama, {\em ibid.\/}, {\bf 60}, 2351 (1991).
\bibitem{Yasuoka1997}H. Yasuoka, Hyperfine Interactions, {\bf 105}, 
27 (1997).
\bibitem{Fujiyama1997}S. Fujiyama, Y. Itoh, H. Yasuoka and Y. Ueda, 
J. 
Phys. Soc. Jpn. {\bf 66}, 2864 (1997).
\bibitem{Itoh1997}Y. Itoh, M. Matsumura and H. Yamagata, J. Phys. 
Soc. Jpn. {\bf 66}, 3383 (1997).
\bibitem{Pennington1991}C. H. Pennington and C. P. Slichter, Phys. 
Rev. 
Lett. {\bf 66}, 381 (1991).
\bibitem{Takigawa1994}M. Takigawa, Phys. Rev. B {\bf 49}, 4158 (1994).
\bibitem{Thelen1994}D. Thelen and D. Pines, Phys. Rev. B {\bf 49}, 
3528 (1994).
\bibitem{Pennington1989}C. H. Pennington, D. J. Durand, C. P. 
Slichter, J. P. Rice, 
E. D. Bukowski and D. M. Ginsberg, Phys. Rev. B {\bf 39}, 274 (1989).
\bibitem{Itoh1992}Y. Itoh, H. Yasuoka, Y. Fujiwara, Y. Ueda, T. 
Machi, 
I. Tomeno, K. Tai, N. Koshizuka and S. Tanaka, J. Phys. Soc. Jpn. 
{\bf 
61}, 1287 (1992).
\bibitem{Hahn1950}E. L. Hahn, Phys. Rev. {\bf 80}, 580 (1950).
\bibitem{Weger}M. Weger, Ph. D. Thesis, University of California (1961).
\bibitem{Klauder1962}J. R. Klauder and P. W. Anderson, Phys. Rev. 
{\bf 125}, 912 (1962).
\bibitem{Slichter}C. P. Slichter, {\em Principles of Magnetic 
Resonance\/}, 
(Springer-Verlag, Berlin, 1990).
\bibitem{Portis1956}A. M. Portis, Phys. Rev. {\bf 104}, 584 (1956).
\bibitem{Recchia}C. H. Recchia, K. Gorny and C. H. Pennington, Phys. 
Rev. B {\bf 54}, 4207 (1996).
\bibitem{Walstedt1995}R. E. Walstedt and S-W. Cheong, Phys. Rev. B 
{\bf 51}, 3163 (1995).
\bibitem{Walstedt1996}R. E. Walstedt and S-W. Cheong, Phys. Rev. B 
{\bf 53}, 6030 (1996). 
\bibitem{Sokol1993}A. Sokol and D. Pines, Phys. Rev. Lett. {\bf 71}, 
2813 (1993).
\bibitem{Chubukov1993}A. V. Chubukov and S. Sachdev, Phys. Rev. Lett. 
{\bf 71}, 169 (1993).
\bibitem{Imai1993}T. Imai, C. P. Slichter, A. P. Paulikas and B. 
Veal, Phys. Rev. B {\bf 47}, 9158 (1993).
\bibitem{Keren1997}A. Keren, H. Alloul, P. Mendels and Y. Yoshinari, 
Phys. Rev. Lett. {\bf 78}, 3547 (1997).
\bibitem{Tranquada1995}J. M. Tranquada, B. J. Sternlieb, J. D. Axe, 
Y. Nakamura and S. Uchida, Nature {\bf 375}, 561 (1995).
\bibitem{Recchia1996}C. H. Recchia, J. A. Martindale, C. H. 
Pennington, W. L. Hults and J. L. Smith, Phys. Rev. Lett.  {\bf 
78}, 3543 (1997).
\end{references}
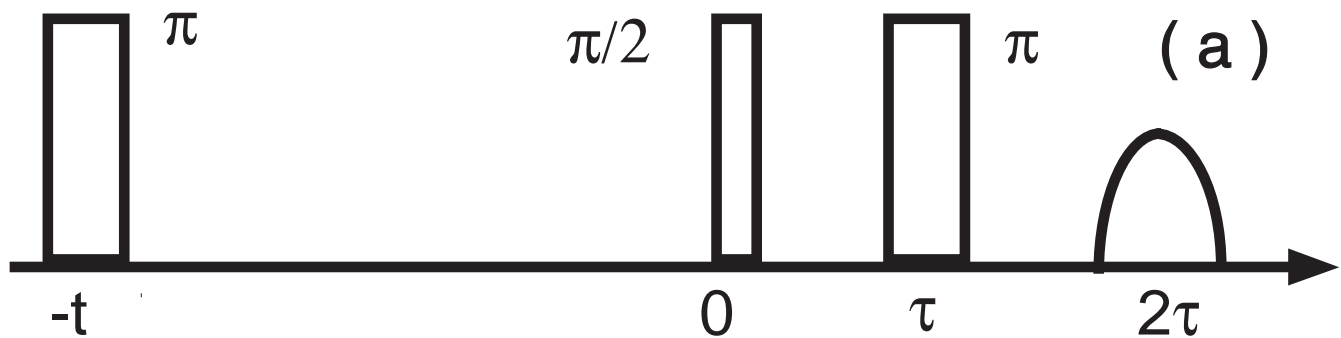
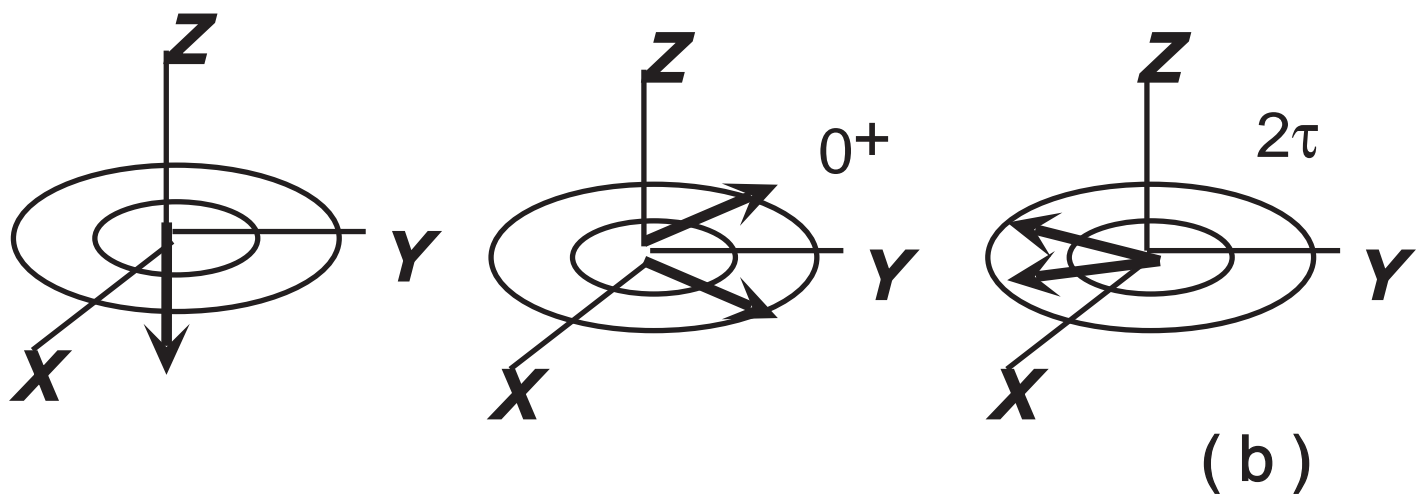
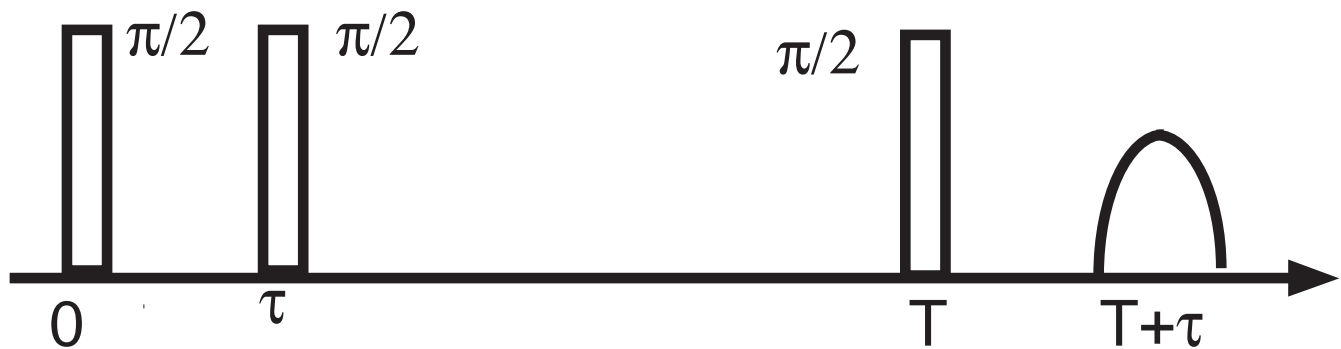
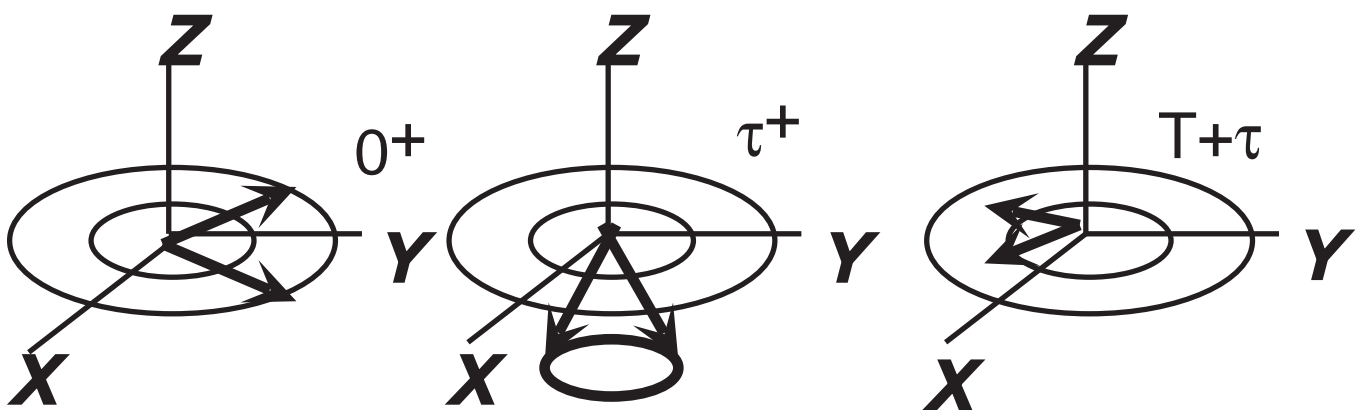
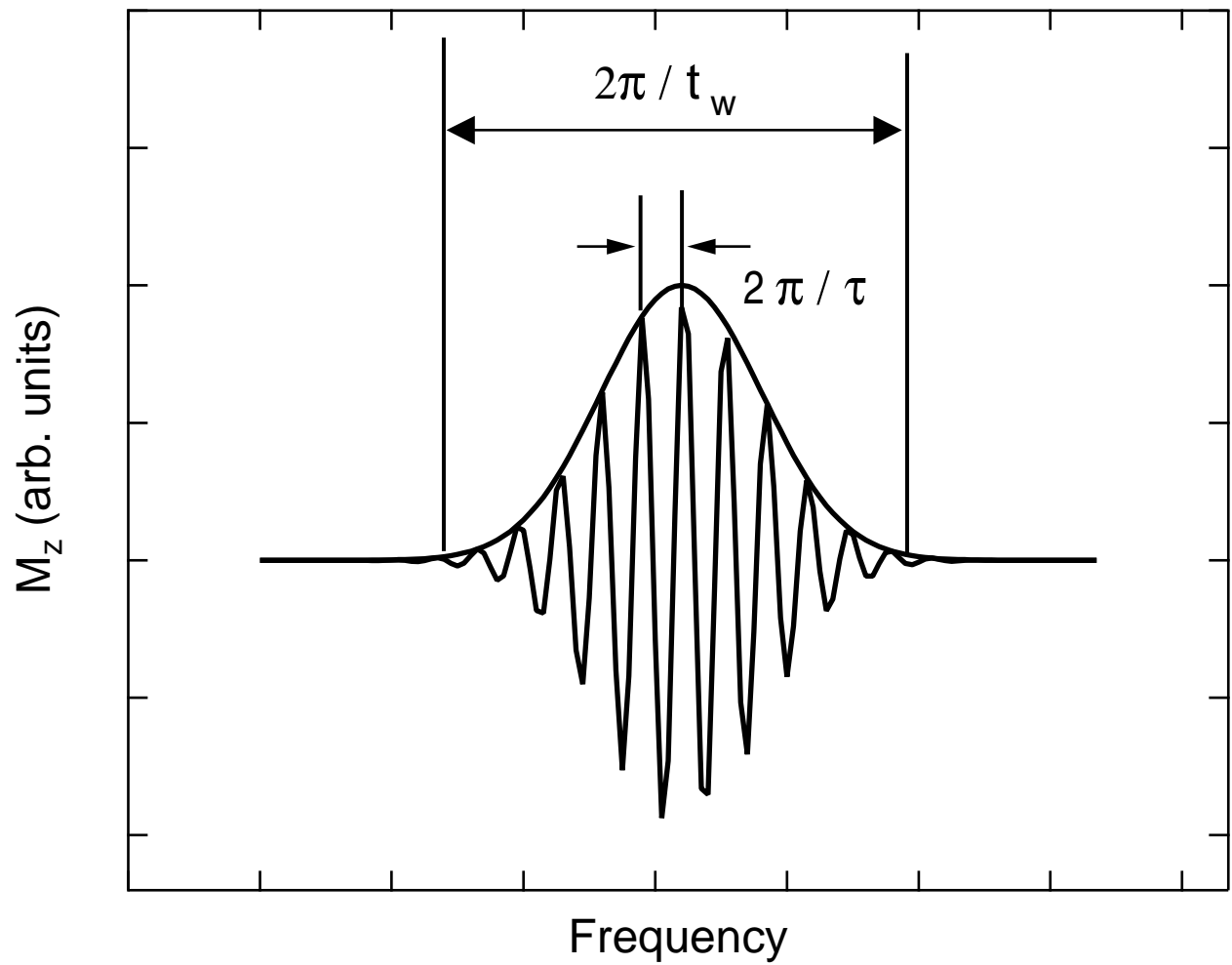
\end{document}